# A coherence method generating macroscopic quantum features using polarization-basis control and its projection measurements of laser light

Byoung S. Ham[1,2,*]
[1]School of Electrical Engineering and Computer Science, Gwangju Institute of Science and Technology, Gwangju 61005, South Korea
[2]Qu-Lidar, Gwangju 61005, South Korea
(Oct. 02, 2024; *bham@gist.ac.kr)

**Abstract**
Quantum entanglement between paired photons is the foundation of optical quantum computing, quantum sensing, and quantum networks. Traditionally, quantum information science has focused on the particle nature of photons at the microscopic scale, often neglecting the phase information of single photons, even for the bipartite quantum entanglement. Recently, a coherence-based approach has been explored to understand the so-called quantum mystery of nonlocal intensity fringes emerging from local randomness. Here, a pure coherence method is presented to create macroscopic quantum features using conventional laser light via linear optics-based measurement modifications. To achieve this, a polarization-basis control of the laser light is conducted to generate indistinguishable characteristics between orthogonally polarized light pairs. Using projection measurements of the polarization-controlled light pairs, we derive coherence solutions of local randomness and nonlocal correlations between independently controlled local parameters, where a fixed relative phase relationship between paired lights is an essential condition to determine the corresponding Bell states.

**Introduction**

Quantum superposition is a fundamental property of quantum mechanics [1,2]. Over the last several decades, quantum correlations between bipartite quantum particles have been intensively studied for quantum superposition of paired photons in terms of product bases focused on the particle nature in quantum mechanics [3-11]. Thus, the phase information of individual photons and between them has been neglected according to the uncertainty principle [12]. Using typical entangled photon pairs generated from spontaneous parametric down-conversion (SPDC) processes [13], polarization-basis manipulations play an important role in demonstrations of local randomness and nonlocal correlations [6-8]. For example, in a cross-sandwiched type I BBO crystal pair, a diagonally polarized pump photon generates superposed entangled photon pairs at an equal probability amplitude of orthogonal polarizations [7]. In ref. 7, the quantum features were experimentally demonstrated for nonlocal intensity fringes between two parties using polarization-basis manipulations, where each party satisfies local randomness. A coherence approach has recently been adapted to ref. 7 to explore the mysterious quantum features, where an inherently fixed phase between paired photons is critical [14]. The local randomness observed [7] has also been coherently examined for the direct result of incoherence between photon pairs generated from different BBO crystals based on the violation of quantum erasers [15-19]. The quantum eraser is for a retrospective quantum feature whose preset particle nature with orthogonal polarization bases turns out to be the wave nature by a post-measurement [18,19], where the wave and particle natures are mutually exclusive in quantum mechanics [2].

Contrary to the expected interference fringes, as in quantum erasers [18,19], typical SPDC-generated entangled photon pairs result in a uniform intensity measured in individual detectors. This is the local randomness in quantum information. The weirdness of local randomness is due to the two-photon intensity fringes when coincidently measured between individual detectors. Recently, the origin of the local randomness has been coherently analyzed for the incoherence feature between entangled photon pairs generated from different crystals rather than for their orthogonal polarization bases [14]. On the other hand, the observed two-photon intensity fringes have been analyzed for the selective measurements of product bases through coincidence detection via polarization-basis manipulations [14]. Interestingly, this selective measurement process requires a fixed phase relationship between entangled photons generated from each BBO crystal [14].

Similarly, such local randomness and nonlocal correlations in ref. 7 have also been demonstrated using type II SPDC for unidirectionally pumped double-slit BBO [11] and bidirectionally pumped single BBO [8] configurations, where corresponding coherence analyses have been followed, respectively [20,21]. Here, a counterintuitive coherence method is presented for creating macroscopic nonlocal correlations using a commercial laser via polarization-basis manipulations. For this, the same setup of linear optics as in ref. 7 is adapted. Selective measurements of polarization-controlled light pairs are carried out for a joint-intensity product, similar to the coincidence detection of single photons, but with photodetectors and analog/digital signal processing. Unlike ref. 7, the preparation of incoherence features in all individual detectors for local randomness is achieved by a nonlinear optic device, whose generated light pulses are orthogonal and timely separated [18].

**Methods**

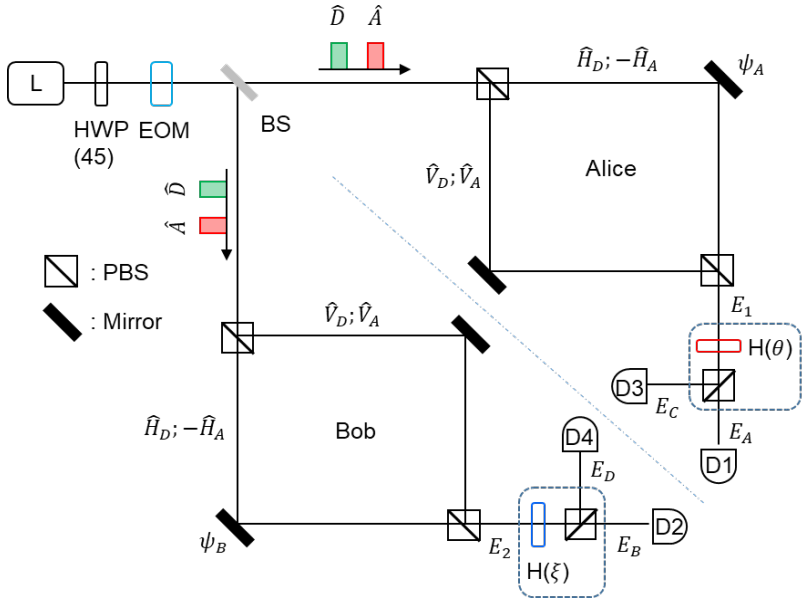

**Fig. 1**. Schematic of macroscopic entanglement generation using polarization-basis manipulation of laser light.

Figure 1 shows the schematic of the proposed macroscopic quantum entanglement generation using laser light via polarization-basis manipulations and their projection measurements. A 22.5-degree rotated half-wave plate (45 HWP) is placed right after the laser L to provide random polarization bases of the lights in horizontal ($\hat{H}$) and vertical ($\hat{V}$) directions. An electro-optic modulator (EOM) follows the 45 HWP to generate orthogonally polarized light pulses in diagonal ($\hat{D}$) and antidiagonal ($\hat{A}$) directions, where the switching time of EOM is set random at a 50% duty cycle. Due to the macroscopic feature of the light pulses, the beam splitter (BS) results in an equal distribution of the polarization-basis superposition state of lights into two parties, Alice and Bob. For the post-measurement control as in quantum erasers [18,19], the linear optics setup composed of a HWP and a polarizing beam splitter (PBS) is inserted into each MZI output port in each party. Here, the role of PBS is for the polarization-basis projection onto its polarization axes. Thus, the polarization-basis projected light pulses are individually and jointly measured for the test of local randomness and nonlocal correlations, respectively. In a space-time separation between two parties, this quantum correlation satisfies the nonlocal quantum feature satisfying violation of Bell's inequality (see Analysis) [6-10].

**Analysis**

For the coherence analysis of the proposed macroscopic quantum entanglement in Fig. 1, the orthogonal polarization bases ($\hat{D}$; $\hat{A}$) of laser light pulses provided by the EOM are manipulated by the setup of linear optics composed of HWP and PBS in each MZI for the indistinguishability of the originally orthogonally polarized lights. The amplitude $E_0$ of the vertically polarized light L is represented by a simple harmonic oscillator of Maxwell's equation. By the 45 HWP, the given polarization direction of the laser light is rotated into the diagonal direction $\hat{D}$. By EOM, the $\hat{D}$-polarized lights are converted into an anti-diagonal one whenever $V_\pi$ voltage is applied. Thus, both $\hat{D}$- and $\hat{A}$-polarized light pulses are prepared in a timely separated manner, satisfying the statistical ensemble of orthogonal bases for the MZI. With random switching of EOM, thus, the polarization state of the lights is also random in the time domain, resulting in a superposition state $E_s = E_0(\hat{D} + \hat{A}e^{i\eta})$, where $\eta$ is a fixed phase induced by EOM.

By the first PBS of MZIs in Fig. 1, the polarization-superposed state $E_s$ is decomposed into $\hat{H}$ and $\hat{V}$, resulting in indistinguishability between $\hat{D}$ and $\hat{A}$. By the second PBS of MZIs, the following relations result in both parties:

$$E_1 = E_0[(\hat{H}_D^\alpha + \hat{V}_D^\alpha) + (-\hat{H}_A^\alpha + \hat{V}_A^\alpha)e^{i\eta_\alpha}]/\sqrt{2}, \tag{1}$$

$$E_2 = E_0[(\hat{H}_D^\beta + \hat{V}_D^\beta) + (-\hat{H}_A^\beta + \hat{V}_A^\beta)e^{i\eta_\beta}]/\sqrt{2}, \tag{2}$$

where the superscript $\alpha$ and $\beta$ represent Alice and Bob, respectively. The subscript D and A represent diagonal and anti-diagonal polarization directions to indicate its original light source, respectively. Additionally, added phases are defined by $\eta_\alpha = \eta + \psi_A$ and $\eta_\beta = \eta + \psi_B$, where $\psi_A$ and $\psi_B$ are local parameters independently controllable with MZI path-length variations.

By the linear optics of HWP and PBS (see the dotted boxes) followed by MZIs, the same polarization basis of the otherwise orthogonally polarized lights is projected onto $\hat{H}$ and $\hat{V}$ axes, resulting in the final measurable events of lights $E_A$, $E_B$, $E_C$, and $E_D$ in amplitudes. Thus, the corresponding intensities are as follows:

$$I_A = I_0[(\hat{H}_D^\alpha \cos\theta + \hat{V}_D^\alpha \sin\theta) + (-\hat{H}_A^\alpha \cos\theta + \hat{V}_A^\alpha \sin\theta)e^{i\eta_\alpha}][(\hat{H}_D^\alpha \cos\theta + \hat{V}_D^\alpha \sin\theta) + (-\hat{H}_A^\alpha \cos\theta + \hat{V}_A^\alpha \sin\theta)e^{-i\eta_\alpha}]/2$$
$$= I_0[(\hat{H}_D^\alpha \hat{H}_D^\alpha \cos^2\theta + \hat{V}_D^\alpha \hat{V}_D^\alpha \sin^2\theta + \hat{H}_A^\alpha \hat{H}_A^\alpha \cos^2\theta + \hat{V}_A^\alpha \hat{V}_A^\alpha \sin^2\theta)]/2$$
$$= I_0, \tag{3}$$

$$I_B = I_0[(\hat{H}_D^\beta \cos\xi + \hat{V}_D^\beta \sin\xi) + (-\hat{H}_A^\beta \cos\xi + \hat{V}_A^\beta \sin\xi)e^{i\eta_\beta}][(\hat{H}_D^\beta \cos\xi + \hat{V}_D^\beta \sin\xi) + (-\hat{H}_A^\beta \cos\xi + \hat{V}_A^\beta \sin\xi)e^{-i\eta_\beta}]/2$$
$$= I_0[(\hat{H}_D^\beta \hat{H}_D^\beta \cos^2\xi + \hat{V}_D^\beta \hat{V}_D^\beta \sin^2\xi + \hat{H}_A^\beta \hat{H}_A^\beta \cos^2\xi + \hat{V}_A^\beta \hat{V}_A^\beta \sin^2\xi)]/2$$
$$= I_0, \tag{4}$$

where, $E_A = E_0[(\hat{H}_D^\alpha \cos\theta + \hat{V}_D^\alpha \sin\theta) + (-\hat{H}_A^\alpha \cos\theta + \hat{V}_A^\alpha \sin\theta)e^{i\eta_\alpha}]/\sqrt{2}$ and $E_B = E_0[(\hat{H}_D^\beta \cos\xi + \hat{V}_D^\beta \sin\xi) + (-\hat{H}_A^\beta \cos\xi + \hat{V}_A^\beta \sin\xi)e^{i\eta_\beta}]/\sqrt{2}$. $\hat{H}_D^\alpha \hat{H}_A^\alpha = \hat{V}_D^\alpha \hat{V}_A^\alpha = \hat{H}_D^\beta \hat{H}_A^\beta = \hat{V}_D^\beta \hat{V}_A^\beta = 0$ is satisfied not by the orthogonal polarization bases but by the no temporal overlap, resulting in incoherence feature between original superposition bases $\hat{H}$ and $\hat{V}$. Likewise, $I_C = I_D = I_0$ is obtained (see Supplementary Materials). Thus, the coherence solutions of the macroscopic local randomness in each party are analytically derived for Fig. 1. These coherence solutions of local randomness imply a single-shot measurement, as in coherence optics.

The joint-intensity correlation between two output fields of Eqs. (3) and (4) from both MZIs is calculated as follows for $\eta_{\alpha\beta} = 2n\pi$ (n = 0, ±1, ±2):

$$R_{AB}(0) = I_0^2[(\hat{H}_D^\alpha \cos\theta + \hat{V}_D^\alpha \sin\theta) + (-\hat{H}_A^\alpha \cos\theta + \hat{V}_A^\alpha \sin\theta)e^{i\eta_\alpha}][(\hat{H}_D^\beta \cos\xi + \hat{V}_D^\beta \sin\xi) + (-\hat{H}_A^\beta \cos\xi + \hat{V}_A^\beta \sin\xi)e^{i\eta_\beta}](cc)/4$$
$$= I_0^2[(\hat{H}_D^\alpha \hat{H}_D^\beta \cos(\theta - \xi) + \hat{H}_D^\alpha \hat{V}_D^\beta \sin(\theta + \xi)) + (\hat{H}_A^\alpha \hat{H}_A^\beta \cos(\theta - \xi) - \hat{H}_A^\alpha \hat{V}_A^\beta \sin(\theta + \xi))e^{i\eta_{\alpha\beta}}](cc)/4$$
$$= I_0^2 \cos^2(\theta - \xi), \tag{5}$$

where $\eta_{\alpha\beta} = \eta_\alpha + \eta_\beta$, $\hat{H}_D^\alpha \hat{V}_D^\beta = \hat{H}_A^\alpha \hat{V}_A^\beta$ due to 50/50 BS, and $\hat{H}_D^\alpha \hat{H}_A^\beta = \hat{V}_D^\alpha \hat{V}_A^\beta = \hat{H}_D^\beta \hat{H}_A^\alpha = \hat{V}_D^\beta \hat{V}_A^\alpha = 0$ due to the no temporal overlap caused by EOM switching. If $\eta_{\alpha\beta} = (2n+1)\pi$, $R_{AB}(0) = I_0^2 \sin^2(\theta + \xi)$ is resulted. These quantum features represent two out of four Bell states: $|\phi^+\rangle$ and $\psi^+\rangle$ [6]. Thus, the critical condition of a fixed phase $\eta_{\alpha\beta}$ is coherently derived for the quantum feature for Fig. 1.

Similarly, $R_{CD} = R_{AB}$ and $R_{AD}(0) = R_{BC}(0) = I_0^2 \sin^2(\theta - \xi)$ are obtained for $\eta_{\alpha\beta} = 2n\pi$ (see Supplementary Materials). If $\eta_{\alpha\beta} = (2n+1)\pi$, $R_{AD}(0) = R_{BC}(0) = I_0^2 \cos^2(\theta + \xi)$ is obtained (see Supplementary Materials). These are other two Bell states: $|\psi^-\rangle$ and $\phi^-\rangle$ [6]. Thus, the typical form of nonlocal correlations satisfying four Bell states in the particle-nature-based quantum mechanics [6] are coherently and macroscopically derived from the joint-intensity measurements of independently controlled output fields between (space-time) separated parties in Fig. 1. More importantly, the contribution of the phase information $\eta_{\alpha\beta}$ to the (nonlocal) quantum feature is not for the absolute distance but for a relative distance between two parties.

For these joint-intensity correlations $R_{AB}$ and $R_{AD}$, no longer a single-photon detector or single-photon counting module is needed. Instead, a digital signal processing unit based on analog-to-digital converters plays a key role in the selective measurements. Most importantly, the coherently derived quantum feature in Eq. (5) is due to the selective measurements caused by the polarization-basis manipulation and its projection onto PBS to induce a complete incoherence feature between orthogonal polarization bases, i.e., $\hat{H}_D^\alpha \hat{H}_A^\beta = \hat{V}_D^\alpha \hat{V}_A^\beta = \hat{H}_D^\beta \hat{H}_A^\alpha = \hat{V}_D^\beta \hat{V}_A^\alpha = 0$. Like the coincidence detection-caused selective measurements [14] between SPDC-generated entangled photons in each pair [7,8], the nonlocal correlation in Eq. (5) is obtained from the temporal overlap caused by polarization-basis manipulations and projection measurements by the same set of linear optics. The price for this nonlocal quantum feature is 50% event loss, as in the conventional method in a microscopic regime [14]. Thus, the selective measurement-caused nonlocal correlation in Eq. (5) is similar to the spectral filtering process, as in the rainbow of sunlight. In this context, the nonlocal quantum features based on local randomness are coherently understood as polarization-basis manipulations of superposed light pulses for selective measurements.

**Discussion**

As derived in Eq. (5), the nonlocal correlation is due to the selective measurements, blocking 50% of unwanted product bases from the product-basis events. This measurement technique is the same as the coincidence detection for SPDC-generated entangled photons. For the proposed macroscopic nonlocal correlations, this selective measurement was provided by EOM-caused diagonally ($\hat{D}$) and anti-diagonally ($\hat{A}$) polarized input light pulses, whose orthogonally polarized light-based product-basis terms between two parties are eliminated due to the temporal separation. Thus, the projection-caused indistinguishable characteristics of the orthogonally polarized lights from $\hat{D}$ or $\hat{A}$ violate the typical quantum-eraser effect due simply to the no temporal overlap condition provided by the EOM, resulting in the local randomness. Thus, the intensity-product fringes seem to originate in the incoherence feature of the product bases measured between two parties, but the selective measurements indicate definite coherence between measured light pulses with the fixed phase relationship $\eta_{\alpha\beta}$ in Eq. (5).

**Conclusion**

Coherence solutions of macroscopic nonlocal correlation were derived from linear optics-based polarization-basis manipulations for the projection measurements using conventional laser light. To induce incoherence conditions between indistinguishably projected lights for local randomness, an electro-optic modulator was used for the no-temporal overlap condition between orthogonally polarized light pulses. For the joint-intensity correlation between independently controlled local parameters, the no temporal overlap condition played an essential role in the selective measurement process to eliminate unwanted product-basis terms for the same incoherence feature between orthogonal bases, satisfying the statistical ensemble in quantum mechanics. However, a fixed phase relationship between selective product bases was also derived as an essential condition of the nonlocal quantum features violating Bell's inequality. This work should open the door to macroscopic quantum technologies, especially for quantum sensing areas to overcome severely limited intensity-product orders far less than N=100.

**Acknowledgments**

This work was supported by the Ministry of Science and ICT, S. Korea, under the ITRC support program (IITP 2024-2021-0-01810). BSH also acknowledges a partial financial support from the GIST Research 2024.

**Author Declarations**

**Conflict of interest**

The author is the founder of Qu-Lidar.

**Author Contributions**

BSH solely wrote the manuscript.

**Data Availability**

All data generated or analyzed during this study are included in this published article.